# COMBINED ANALYSIS OF SOLAR NEUTRINO AND SOLAR IRRADIANCE DATA: FURTHER EVIDENCE FOR VARIABILITY OF THE SOLAR NEUTRINO FLUX AND ITS IMPLICATIONS CONCERNING THE SOLAR CORE

P.A. STURROCK

*Center for Space Science Astrophysics,*
*Varian 302, Stanford University,*
*Stanford, CA 94305-4060, U.S.A.*
*(e-mail: sturrock@stanford.edu)*

**Abstract.** A search for any particular feature in any single solar neutrino dataset is unlikely to establish variability of the solar neutrino flux since the count rates are very low. It helps to combine datasets, and in this article we examine data from both the Homestake and GALLEX experiments. These show evidence of modulation with a frequency of 11.85 yr$^{-1}$, which could be indicative of rotational modulation originating in the solar core. We find that precisely the same frequency is prominent in power spectrum analyses of the ACRIM irradiance data for both the Homestake and GALLEX time intervals. These results suggest that the solar core is inhomogeneous and rotates with sidereal frequency 12.85 yr$^{-1}$. We find, by Monte Carlo calculations, that the probability that the neutrino data would by chance match the irradiance data in this way is only 2 parts in 10,000. This rotation rate is significantly lower than that of the inner radiative zone (13.97 yr$^{-1}$) as recently inferred from analysis of Super-Kamiokande data, suggesting that there may be a second, inner tachocline separating the core from the radiative zone. This opens up the possibility that there may be an inner dynamo that could produce a strong internal magnetic field and a second solar cycle.

1. Introduction

There have been claims for variability in several solar neutrino datasets, beginning with early claims (see, for instance, Davis 1996) concerning Homestake (Davis, Harmer, and Hoffman, 1968; Cleveland et al. 1998) data, and more recently with claims (see, for instance, Sturrock 2008a) concerning Super-Kamiokande (Fukuda et al., 2001, 2002, 2003) data. However, the experiment that has so far yielded the strongest evidence for variability is the GALLEX experiment (Anselmann *et al*., 1993, 1995; Hampel *et al.,* 1996, 1999), which was in operation from 1991.367 to 1997.060. ). (For convenience in data analysis, all dates are given in "neutrino years" (Sturrock 2004) which start from 1970.000 and run uniformly forward, each year having a duration of 365.2564 days.)

The GALLEX experiment yields three separate indicators of variability: (a) there is strong evidence that the histogram formed from flux measurements is bimodal (Sturrock 2008b), whereas a constant flux should yield a unimodal distribution; (b) there is evidence of a secular change in flux over a time-scale of a year or so (Sturrock, Caldwell, and Scargle, 2006); and (c) there is evidence of periodic modulation at frequency 11.85 $yr^{-1}$ that is suggestive of rotational modulation, and was noted also in analysis of Homestake data (Sturrock, 2008c). These results point to a sidereal rotation rate of 12.85 $yr^{-1}$, which is lower than current estimates of the rotation rate of the radiative zone, if it is assumed to have a uniform rotation rate (Schou *et al*. 1998; Garcia *et al*., 2008). However, helioseismology data do not at present yield well-defined estimates of the rotation rate in or near the solar core (where nuclear burning occurs) that has an outer normalized radius of about 0.2 .

There are therefore at least two reasons to search for independent evidence of solar variability that might be related to evidence found in solar neutrino data: (a) such evidence may confirm the reality of the neutrino variability; and (b) such evidence, in combination with neutrino evidence, may offer new insight into processes in or near the solar core. It appears (surprising as it may seem) that solar irradiance measurements offer such independent evidence.

In Section 2 we analyze Homestake data, that starts at 1970.281, but we truncate the dataset at run 115 that ends at 1991.265, so as to avoid overlap with the GALLEX experiment. The omitted runs 116 through 133 (a time interval 1991.265 to 1994.388) represent only 13 percent of the total Homestake dataset. We also carry out a power-spectrum analysis of ACRIM data (see, for instance, Willson 1979, 2001, and www.acrim.com), kindly compiled and made available by Judith Lean, from the start of that experiment (1978.877) up to the beginning of the GALLEX experiment (1991.367). We then search for a peak that is common to both power spectra by means of a procedure (the joint power statistic) recently developed for this purpose (Sturrock, Scargle, Walther, and Wheatland, 2005). In Section 3, we carry out similar calculations for the GALLEX experiment.

In Section 4, we carry out a comparative analysis of all four power spectra, and we carry out a Monte Carlo calculation to evaluate the probability that the neutrino data happen by chance to yield peaks in their power spectra that correspond to a notable and consistent peak in the ACRIM power spectra. The results are reviewed and discussed in Section 5.

## 2. Power-Spectrum Analyses for the Homestake Interval

As explained in Section 1, we here analyze Homestake data (Davis, Harmer, and Hoffman, 1968; Cleveland et al. 1998) from its beginning in 1970.281 until the beginning of the GALLEX experiment in 1991.367. This leaves 98 runs for analysis. We carry out a power-spectrum analysis using a likelihood procedure that takes account of the start time and end time of each run and the flux estimate of each run (Sturrock, 2003). For consistency, we adopt the standard deviation of the flux estimates as the error term (as in the Lomb-Scargle procedure: Lomb, 1976; Scargle, 1982) for all power-spectrum analyses in this article. The resulting power spectrum, for the frequency band $0 – 20$ $yr^{-1}$, is shown in Figure 1, and the top ten peaks are listed in Table 1. We see that the second strongest peak in this band is one at frequency 11.84 yr$^{-1}$ with power S = 5.90.

We next analyze the ACRIM irradiance measurements for the same time interval. Since irradiance measurements are strongly modulated by the solar cycle, we first subtract 600-day running means to get rid of that modulation. Since data are missing for some days, we use the Lomb-Scargle procedure (Lomb, 1976; Scargle, 1982) for power-spectrum analysis. The resulting power spectrum, for the frequency band 0 – 20 yr$^{-1}$, is shown in Figure 2, and the top twenty peaks are listed in Table 2. We see that there is a peak at frequency 11.85 yr$^{-1}$, with power S = 21.67, essentially the same frequency as that of the second peak in the Homestake power spectrum listed in Table 1.

A convenient procedure for searching for common features in two or more power spectra is to form the "joint power statistic" (Sturrock, Scargle, Walther, and Wheatland, 2005). For just two power spectra, this is given to good approximation by the formula

$$J = \frac{1.94\, Y^2}{0.65 + Y}, \qquad (1)$$

where

$$Y = (S_1 * S_2)^{1/2}. \qquad (2)$$

This statistic has the property that if each power spectrum is distributed exponentially (so that the probability of obtaining the power S or more is $e^{-S}$), then the same is true of the statistic.

We may apply this procedure to the power spectra derived from Homestake and ACRIM data (for the Homestake time interval) to find out whether there are correspondences other than the one noted above. The result is shown in Figure 3, and the top ten peaks are listed in Table 3. We see that the biggest peak is indeed that at frequency 11.84 yr$^{-1}$, for which J = 20.04.

### 3. Power-Spectrum Analyses for the GALLEX Interval

We now carry out a similar analysis of GALLEX data and of ACRIM data for the GALLEX time interval. A likelihood analysis of GALLEX data, similar to the preceding analysis of Homestake data, yields the power spectrum shown in Figure 4, and the top ten peaks are listed in Table 4. [One may note that a recent time-frequency analysis of GALLEX data (Sturrock, 2008c) indicates that the principal modulation is that at 11.87 $yr^{-1}$, of which the peak at 13.64 $yr^{-1}$ is an alias caused by the regularity of the experimental schedule. (Runs are typically spaced at intervals of three or four weeks.)]

A Lomb-Scargle analysis of ACRIM data for the GALLEX interval (1991.367 to 1997.060) leads to the power spectrum shown in Figure 5, and the top ten peaks are listed in Table 5. The biggest peak has a frequency close to 1 $yr^{-1}$. Claus Frohlich and Judith Lean (private communications) have pointed out that this peak is most likely attributable to the Earth's orbit, since the typical north-south asymmetry in the sunspot distribution will lead to an annual variation in the measured irradiance.

The second biggest peak in the range 0 – 20 $yr^{-1}$ is found at 11.87 $yr^{-1}$. The joint power statistic is shown in Figure 6, and the top ten peaks are listed in Table 6. We see that the strongest common feature of the two power spectra is found at 11.87 $yr^{-1}$.

### 4. Combined Analysis of Homestake, GALLEX, and ACRIM data

It is interesting to form the joint power statistic for other combinations of the power spectra. We show in Figure 7 the statistic formed from the Homestake power spectrum and the GALLEX power spectrum, and we list in Table 7 the top ten peaks. We see that the principal peak is at 11.84 $yr^{-1}$.

We show in Figure 8 the statistic formed from the ACRIM power spectrum for the Homestake interval and the ACRIM power spectrum for the GALLEX interval. and we list the top ten peaks in Table 8. We see that, after two very strong peaks near 1 yr-1, that are presumably due to orbital effects, we find a pair of peaks at 11.85 yr$^{-1}$ and at 11.96 yr$^{-1}$, both with values close to J = 50. We see that modulation at or near 11.85 yr$^{-1}$ is present in all four datasets.

We also show the joint power statistic that may be formed from all four power spectra. This statistic may be computed to good approximation (Sturrock, Scargle, Walther, and Wheatland, 2005) from

$$J = \frac{3.88\, Y^2}{1.27 + Y}, \qquad (3)$$

where

$$Y = (S_1 * S_2 * S_3 * S_4)^{1/4}. \qquad (4)$$

The result is shown in Figure 9, and the top ten peaks are listed in Table 9. The peak at 0.92 yr$^{-1}$ no longer appears, showing that it is a property of irradiance data but not of neutrino data. On the other hand, there is a strong peak at 11.84 yr$^{-1}$, with $J \approx 41$, confirming that this modulation occurs consistently in both irradiance and neutrino data. Apart from the nearby peak at 11.93 yr$^{-1}$, with $J \approx 21$, the next biggest peak is one at 6.04 yr$^{-1}$, with $J \approx 19$.

Finally, we pose the question: What is the chance of finding in both neutrino datasets modulation that happens to agree with a notable modulation in irradiance data? We address this question by carrying out Monte Carlo calculations. For each simulation, we shuffle the Homestake data and (separately) the GALLEX data. For each shuffle, we keep the flux value and the duration of each run together, but we re-assign this pair of quantities among the list of end-times. For each simulation, we compute the maximum value of the fourth-order joint power statistic. We restrict these calculations to the frequency range 10 – 15 yr$^{-1}$, which is wide enough to encompass all relevant possible values of the synodic rotation frequency. (However, the result is insensitive to the range selected.) The result is shown in Figure 10. For only 20 simulations out of 100,000 do we find a value of the JPS that is as large as or larger than the value (40.87) derived from the actual data. Hence the case for rotational modulation of both Homestake and GALLEX data at a frequency found in ACRIM irradiance data may be accepted at the 99.98% confidence level.

## 5. Discussion

The fact that Homestake and GALLEX data show modulation at a frequency (11.85 yr$^{-1}$) that shows up prominently in irradiance data is significant for both particle physics and solar physics. For particle physics, it establishes – at a high confidence level (99.98%) - that the solar neutrino flux is variable. For solar physics, it establishes that the solar irradiance is modulated by a process that is also involved in the production and/or propagation of neutrinos. However, these two inferences raise many questions.

If (as this article and other evidence imply) the solar neutrino flux is variable, one must view with suspicion analyses that are based on the assumption that it is constant. If the flux is variable on the timescale of weeks, it may also be variable on the timescale of years, or decades, or longer. One of our analyses (Sturrock, Caldwell, and Scargle, 2004) already points to variation of the GALLEX measurements on a timescale of years. This means that we must be cautious about assuming that the production rate of neutrinos can be inferred from the solar luminosity, since this assumption may or may not prove to be correct.

It will be essential to understand the causes of variability of the neutrino flux. The periodic modulation that is the central feature of this article is presumably due to rotational modulation. The depth of modulation has been found to be quite big (of order 80%, Sturrock 2008c) which seems difficult to reconcile with the RSFP (Resonant Spin Flavor Precession; Akhmedov, 1988; Lim and Marciano, 1988) process (Balantekin and Volpe, 2004; Chauhan and Pulido, 2004; Chauhan, Pulido, and Picariello, 2007; Pulido, Chauhan, and Picariello, 2007). It could perhaps originate in a core with asymmetric nuclear burning, in combination with the MSW process (Mikheyev and Smirnov, 1986; Wolfenstein, 1978). Analysis of this scenario cannot be based on the usual assumptions of spherical symmetry and radial neutrino velocity vectors. On the other hand, recent evidence for r-mode oscillations in Super-Kamiokande data (Sturrock, 2008a) is probably indicative of the RSFP process in the radiative zone, involving of course an asymmetric magnetic-field configuration.

The fact that the solar irradiance exhibits modulation identical to that of neutrinos leads to the rather surprising conclusion that the irradiance is modulated by processes in the solar core. Another surprise is that the core appears to rotate at a slower rate than that of the radiative zone, as inferred from Super-Kamiokande data (Sturrock, 2008a), indicating that the core loses angular momentum by a process other than viscous interaction with the radiative zone. It is possible that both questions may be answered in terms of gravito-acoustic waves related to the "five-minute" oscillations in the Sun. (See, for instance, Unno, Osaki, Ando, and Shibahashi 1979.) These waves grow in amplitude (approximately as the inverse square root of the density) in propagating from the core to the photosphere, so that a very small amplitude in the core could become a significant amplitude at the photosphere, large enough perhaps to show up as a modulation of the irradiance. Such waves can carry angular momentum and thereby exert a decelerating torque on the core, analogous to the deceleration of neutron stars by gravitational radiation (Cutler and Thorne, 2002). This scenario of course requires that one abandon the usual assumption that the core is in a steady, spherically symmetric state. It requires one to consider the possibility that the core is fluctuating and asymmetric.

The fact that the core apparently rotates more slowly than the radiative zone raises the intriguing possibility that there may be a second tachocline, similar to the jump in rotation rate at normalized radius 0.7 that separates the radiative zone from the convection zone. It is accepted that the outer tachocline is the seat of an oscillatory dynamo that is responsible for the 11-year solar cycle (22-years, if one takes account of the reversal in polarity of the magnetic field). (See, for instance, Krause, Raedler, and Ruediger, 1993.) A similar dynamo at an inner tachocline could generate magnetic fields of strength 10 M gauss or more. This raises the question of whether there is any evidence for an oscillation that might be related to such a dynamo. On examining Figure 9 and Table 9 we find that, after the two peaks that we attribute to rotational modulation in the core, the next strongest peak is one at 6.04 yr$^{-1}$. We regard this as a candidate frequency for a second solar cycle, but there are other possibilities that need to be explored.

The existence of an inner tachocline may also offer an explanation for the excitation of r-mode oscillations in the radiative zone, as inferred from Super-Kamiokande data (Sturrock 2008a). This process would be analogous to the excitation of r-modes in the outer tachocline that offers an explanation of the Rieger periodicities (Bai, 1992, 1994; Rieger et al. 1984; Bai and Cliver, 1990; Kile and Cliver, 1991; Rieger et al. 1984).

The apparent discovery of consistent modulation in both solar neutrinos and solar irradiance seems to give a definite and positive answer the question of whether or not the solar neutrino flux is variable, but it seems also to raise new and intriguing questions.

## Acknowledgements
Thanks are due to Bala Balantekin, Alexander Kosovichev, Joao Pulido, and Jeffrey Scargle for helpful discussions related to this work, which was supported by NSF Grant AST-0607572. Special thanks are due to the Claus Frohlich and Judith Lean for making irradiance data available for this analysis.

11/22

# Tables

## TABLE 1

Top ten peaks in the frequency band $0 - 20$ yr$^{-1}$ in the Homestake power spectrum.

| Frequency (yr$^{-1}$) | Power |
|---|---|
| 3.14 | 6.11 |
| 11.84 | 5.90 |
| 8.81 | 5.61 |
| 15.82 | 5.45 |
| 1.84 | 5.06 |
| 3.70 | 4.67 |
| 19.19 | 4.62 |
| 18.10 | 4.61 |
| 9.19 | 4.54 |
| 18.01 | 4.35 |

TABLE 2

Top twenty peaks in the frequency band 10 – 20 yr$^{-1}$ in the power spectrum formed from ACRIM data for the Homestake time interval.

| Frequency (yr$^{-1}$) | Power |
|---|---|
| 0.72 | 49.04 |
| 0.91 | 47.99 |
| 1.10 | 45.11 |
| 1.59 | 32.59 |
| 0.56 | 31.14 |
| 11.96 | 27.04 |
| 7.33 | 26.97 |
| 0.82 | 26.79 |
| 6.84 | 23.08 |
| 1.50 | 22.02 |
| 11.85 | 21.67 |
| 4.29 | 20.27 |
| 8.39 | 19.28 |
| 12.31 | 18.98 |
| 12.07 | 18.55 |
| 2.14 | 17.97 |
| 1.31 | 17.97 |
| 1.40 | 17.29 |
| 15.73 | 16.95 |
| 3.44 | 16.86 |

TABLE 3

Top ten peaks in the frequency band 0 – 20 yr$^{-1}$ in the joint power statistic formed from the power spectra formed from Homestake data and from ACRIM data for the Homestake interval.

| Frequency (yr$^{-1}$) | Power |
|---|---|
| 11.84 | 20.04 |
| 0.74 | 16.45 |
| 1.08 | 16.35 |
| 15.82 | 15.65 |
| 3.14 | 14.54 |
| 0.56 | 12.85 |
| 8.81 | 12.43 |
| 7.32 | 11.77 |
| 0.82 | 11.29 |
| 15.52 | 10.75 |

TABLE 4

Top ten peaks in the frequency band 10 – 20 yr$^{-1}$ in the GALLEX power spectrum.

| Frequency (yr$^{-1}$) | Power |
|---|---|
| 13.64 | 7.81 |
| 13.08 | 6.08 |
| 4.54 | 5.73 |
| 11.87 | 5.51 |
| 6.93 | 5.02 |
| 6.05 | 4.81 |
| 3.97 | 4.62 |
| 12.26 | 3.83 |
| 10.72 | 3.41 |
| 14.02 | 3.15 |

TABLE 5

Top ten peaks in the frequency band 10 – 20 yr$^{-1}$ in the power spectrum formed from ACRIM data for the GALLEX time interval.

| Frequency (yr$^{-1}$) | Power |
|---|---|
| 0.96 | 59.18 |
| 11.87 | 32.31 |
| 5.09 | 29.10 |
| 6.02 | 22.82 |
| 7.35 | 22.10 |
| 3.48 | 20.16 |
| 1.44 | 19.58 |
| 12.38 | 18.09 |
| 1.62 | 17.70 |
| 1.19 | 16.88 |

TABLE 6

Top ten peaks in the frequency band 0 – 20 yr$^{-1}$ in the joint power statistic formed from the power spectra formed from GALLEX data and from ACRIM data for the GALLEX interval.

| Frequency (yr$^{-1}$) | Power |
|---|---|
| 11.87 | 24.13 |
| 6.04 | 19.13 |
| 5.11 | 13.77 |
| 6.91 | 12.37 |
| 4.53 | 10.86 |
| 14.01 | 10.28 |
| 13.62 | 10.01 |
| 6.29 | 9.45 |
| 7.84 | 9.40 |
| 3.53 | 9.12 |

TABLE 7

Top ten peaks in the frequency band 0 – 20 yr$^{-1}$ in the joint power statistic formed from the Homestake and GALLEX power spectra.

| Frequency (yr$^{-1}$) | Power |
|---|---|
| 11.84 | 9.54 |
| 13.61 | 7.08 |
| 13.68 | 5.99 |
| 10.72 | 5.92 |
| 6.08 | 5.79 |
| 11.91 | 4.96 |
| 4.00 | 4.64 |
| 1.84 | 4.62 |
| 0.22 | 4.42 |
| 10.53 | 4.15 |

TABLE 8

Top ten peaks in the frequency band 0 – 20 yr$^{-1}$ in the joint power statistic formed from the ACRIM power spectra for the Homestake and GALLEX time intervals.

| Frequency (yr$^{-1}$) | Power |
|---|---|
| 0.92 | 96.95 |
| 1.08 | 55.90 |
| 11.96 | 50.37 |
| 11.85 | 49.77 |
| 1.15 | 48.60 |
| 7.33 | 45.12 |
| 1.60 | 44.46 |
| 0.83 | 40.91 |
| 1.49 | 35.06 |
| 3.44 | 32.64 |

TABLE 9

Top ten peaks in the frequency band 0 – 20 yr$^{-1}$ in the joint power statistic formed from the Homestake and GALLEX power spectra, together with ACRIM power spectra for the Homestake and GALLEX time intervals.

| Frequency (yr$^{-1}$) | Power |
|---|---|
| 11.85 | 40.87 |
| 11.93 | 21.34 |
| 6.04 | 19.25 |
| 1.08 | 18.64 |
| 0.74 | 17.56 |
| 0.57 | 15.61 |
| 7.32 | 14.81 |
| 0.82 | 13.41 |
| 7.02 | 13.21 |
| 5.13 | 13.11 |

FIGURES

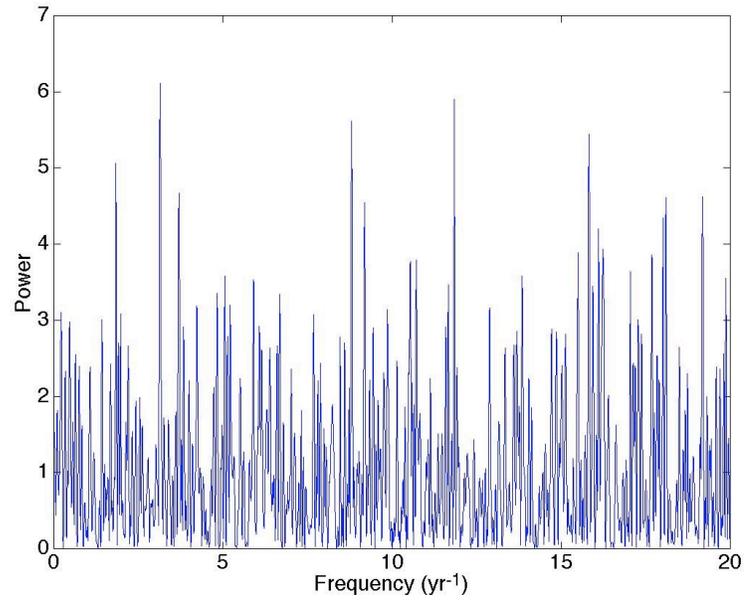

Figure 1. Power Spectrum of Homestake Data, truncated to avoid overlap with GALLEX data.

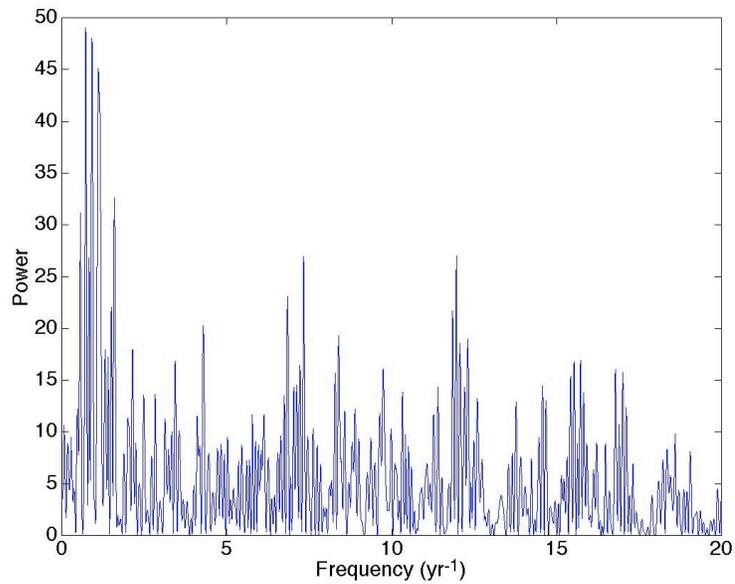

Figure 2. Power Spectrum of detrended ACRIM Data, from the commencement of the experiment up to the beginning of the GALLEX experiment.

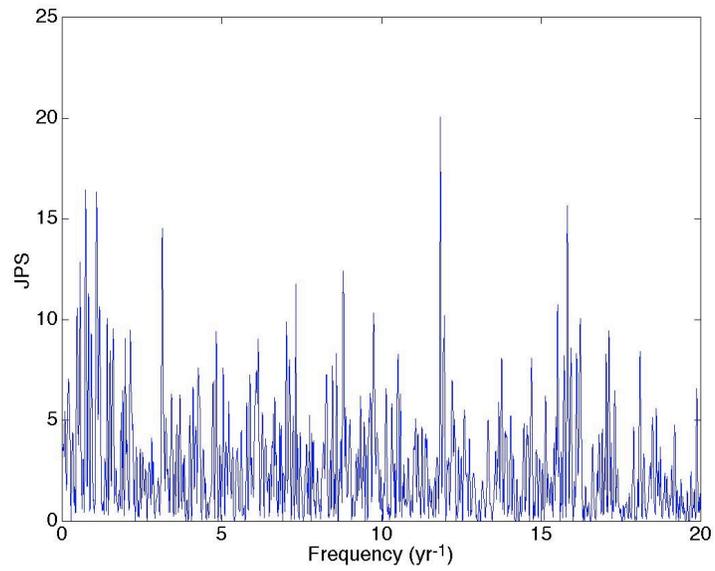

Figure 3. Joint power statistic formed from the Homestake power spectrum and the power spectrum for the Homestake interval of ACRIM data.

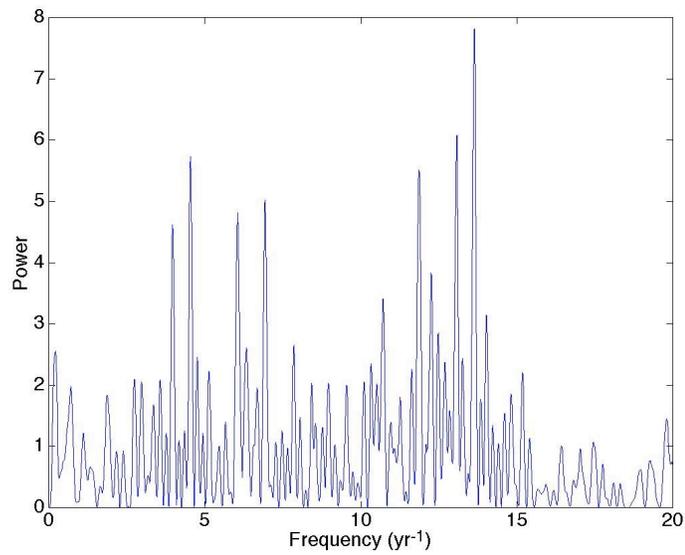

Figure 4. Power Spectrum of GALLEX Data.

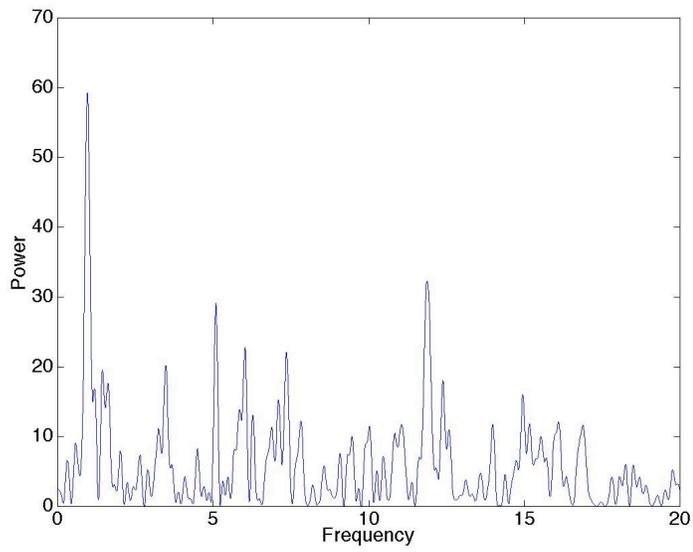

Figure 5. Power Spectrum of detrended ACRIM Data, for the interval of operation of the GALLEX experiment.

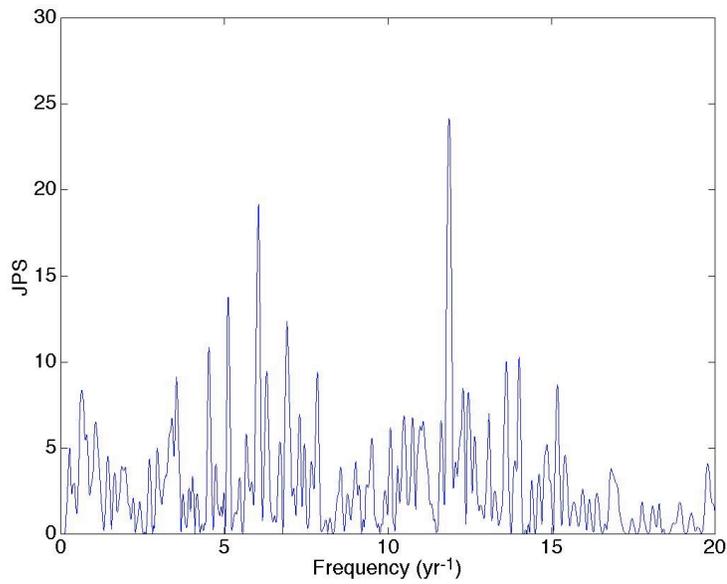

Figure 6. Joint power statistic formed from the GALLEX power spectrum and the power spectrum for the GALLEX interval of ACRIM data.

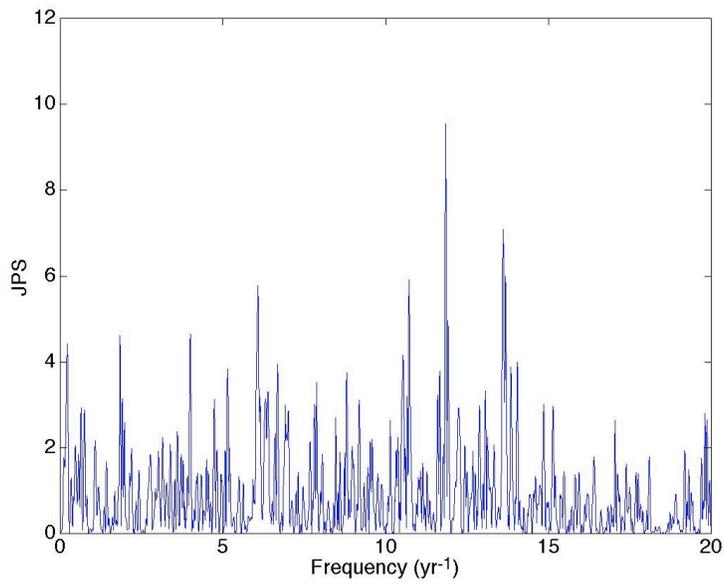

Figure 7. Joint power statistic formed from the Homestake power spectrum and the GALLEX power spectrum.

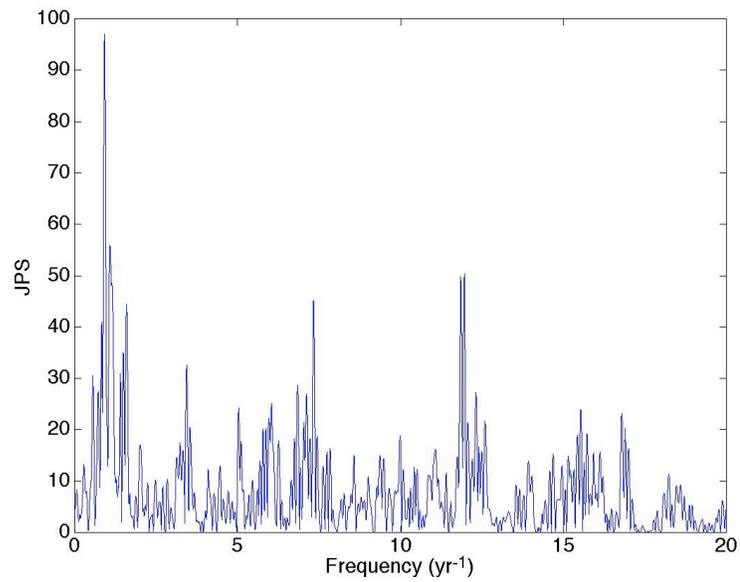

Figure 8. Joint power statistic formed from the ACRIM power spectrum for the Homestake interval and the ACRIM power spectrum for the GALLEX interval.

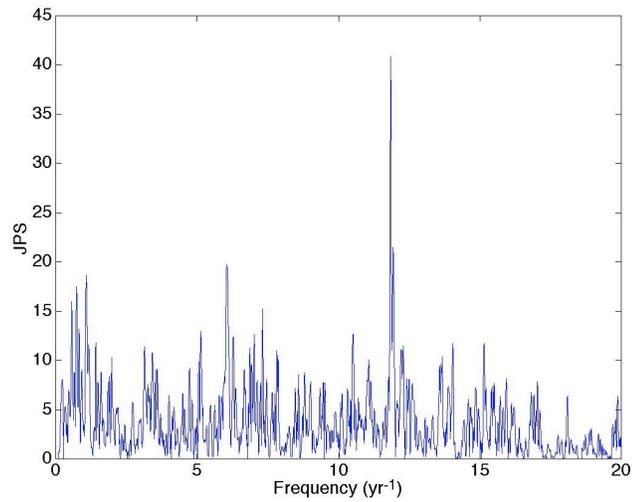

Figure 9. Joint power statistic formed from the Homestake power spectrum, the power spectrum for the Homestake interval of ACRIM data, the GALLEX power spectrum, and the power spectrum for the GALLEX interval of ACRIM data.

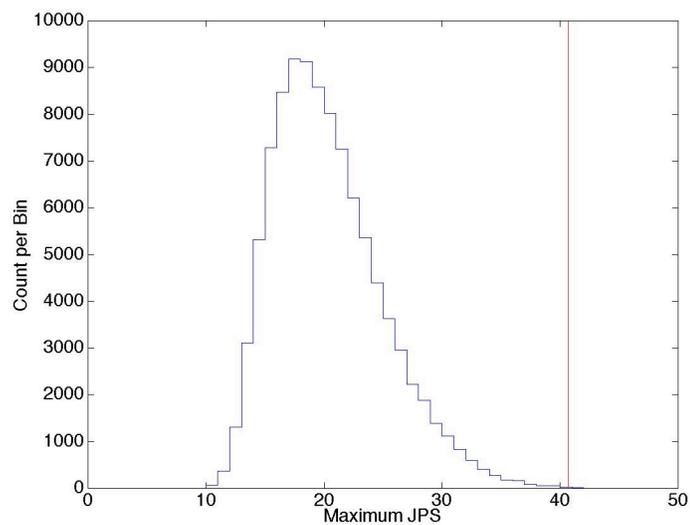

Figure 10. Histogram of results of Monte Carlo simulations, shuffling both Homestake and GALLEX data. Of 100,000 simulations of the calculation shown in Figure 9, only 20 have a joint power statistic larger than the value (40.87) for the actual data.